\documentclass[twocolumn,prd,aps,groupedaddress,nopacs,nofootinbib]{revtex4-1}
\usepackage{amsmath}
\usepackage{amssymb}
\usepackage{latexsym}
\usepackage{graphicx}
\usepackage{hyperref}
\usepackage{mathrsfs}
\usepackage{color}
\usepackage{bm}
\usepackage[caption=false]{subfig}

\begin{document}

\title{Optimising Gaussian processes for reconstructing dark energy dynamics from supernovae}

\author{Marina Seikel$^{1,2}$ and Chris Clarkson$^2$}
\affiliation{ 
\it $^1$Department of Physics, University of the Western Cape, Cape Town 7535, South Africa\\
\it $^2$Astrophysics, Cosmology \& Gravity Centre,
and, Department of Mathematics \& Applied Mathematics,
University of Cape Town,
Cape Town 7701,
South Africa
}

\begin{abstract}
Gaussian processes are a fully Bayesian smoothing technique that allows for the reconstruction of a function and its derivatives directly from
observational data, without assuming a specific model or choosing a
parameterization. This is ideal for constraining dark energy  because physical models are generally phenomenological and poorly motivated. Model-independent constraints on dark energy are an especially important alternative to parameterized models, as the priors involved have an entirely different source so can be used to check constraints formulated from models or parameterizations. A critical prior for Gaussian process reconstruction lies in the choice of covariance function.
We show how the
choice of covariance function affects the result of the
reconstruction, and present a choice which leads to reliable results for present day supernovae data. We also introduce a
method to quantify deviations of a model from the Gaussian process
reconstructions.
\end{abstract}

\maketitle

\section{Introduction}

The $\Lambda$CDM of cosmology has been very successful in describing observations of
various  data. While it is the simplest model that is
consistent with the data, there is a plethora of other viable
cosmological models (for reviews see \cite{Copeland:2006wr,
  Clifton:2011jh}). A common approach is to fit these models to the
data and apply model selection techniques to determine the most likely
theory.

An alternative approach is to perform model-independent
reconstructions, which can be used to determine deviations from
$\Lambda$CDM. This is particularly important when physical models are poorly motivated from a fundamental theory, as is the case for most dark energy models. A range of consistency tests have been introduced to
probe various aspects of $\Lambda$CDM \cite{Clarkson:2007pz,
  Zunckel:2008ti, Sahni:2008xx, Shafieloo:2009hi, Nesseris:2010ep,
  Seikel:2012cs}. For these tests it is crucial to use a
non-parametric method to reconstruct functions such as the luminosity
distance $d_{L}(z)$ or the Hubble rate $H(z)$. Gaussian processes (GP)
provide such a method, which is also capable of reliably
reconstructing derivatives of a function. GP have been applied to this problem in
cosmology in \cite{Seikel:2012uu, Seikel:2012cs, Yahya:2013xma,
  Holsclaw:2010nb, Holsclaw:2010sk, Holsclaw:2011wi, Shafieloo:2012ht, Shafieloo:2012ms, Nair:2013sna}.

In this work, we focus on some aspects of GP reconstructions of dark
energy from supernova (SN) data. The luminosity distance is given by
\begin{equation}
d_{L}(z) = \frac{c(1+z)}{H_0 \sqrt{-\Omega_k}}
\sin{\left( 
\sqrt{-\Omega_k}\int_0^z{\mathrm{d}z'\frac{H_0}{H(z')}}
\right)}\,,
\end{equation}
where
\begin{eqnarray}
&& H(z)^2 =  H_0^2 \left\{ \Omega_{m} (1+z)^3 + \Omega_{k}(1+z)^2 \right. \\
&&\qquad{} + \left. (1 - \Omega_m - \Omega_k)\exp{\left[3\int_0^z
\frac{1+w(z')}{1+z'}\mathrm{d}z'\right]} \right\}\,. \nonumber
\end{eqnarray}
Using the dimensionless comoving distance
\begin{equation}
D(z)=(H_0/c)(1+z)^{-1}d_L(z)\,,
\end{equation}
we can write the equation of state $w(z)$ as
\begin{eqnarray}
&&\!\!\!w(z)=\{2(1+z)(1+\Omega_kD^2)D''-[(1+z)^2\Omega_kD'^2\nonumber\\&&
+2(1+z)\Omega_kDD'-3(1+\Omega_kD^2)]D'\}/ \label{w} \\&&
\{3\{(1+z)^2[\Omega_k+(1+z)\Omega_m]D'^2-(1+\Omega_kD^2)\}D'\}\,,\nonumber
\end{eqnarray}
where $'=d/dz$.

Reconstructing $w(z)$ from distance data requires the knowledge of the
density parameters $\Omega_m$ and $\Omega_k$. As there are
degeneracies between the density parameters and the equation of state
\cite{Kunz:2007rk, Clarkson:2007bc, Hlozek:2008mt, Seikel:2012uu},
these quantities cannot be determined by considering distance
measurements only. In this work, we only use mock data sets and thus
have the luxury of being able to fix the values for the density
parameters. We use $\Omega_m=0.3$ and $\Omega_k=0$ throughout the
paper. In practice, these degeneracies are
problematic.  Despite these concerns, we
focus on the reconstruction of $w(z)$ with fixed density parameters in
this work because our main interest here is to analyze how well the
input model (which is determined by its equation of state) can be
recovered by GP reconstructions. 
While we focus on the reconstruction of $w$, we also reconstruct the
deceleration parameter,
\begin{equation}
q(z) = - (1+z) D''(z)/D'(z) - 1\,,
\end{equation}
for some tests.

For a non-parametric regression method such as GP, our aim is to produce confidence limits such that we trap the true function appropriately. For example, for repeated realisations of a data set for a given model we require that our confidence limit of 95\% traps the correct model in 95\% of realisations. For~$w$  we are not just trying to reconstruct a function from data, but a complicated function of its first and second derivatives as well. This makes it a much more complicated problem for which we need to identify appropriate covariance functions which can reproduce expected models accurately. 
We introduce a family of 4 $w(z)$'s in addition to LCDM of increasing
complexity which we attempt to reconstruct with a fixed number of SNIa
(about 500). These models are not intended to be physical, but rather to
illustrate the types of features we might be able to constrain.  We
analyse 4 different covariance functions which each have one amplitude
parameter and one length parameter. These are a simple Gaussian, and 3
functions of the Mat\'ern class. 
The main difference between these is the widths of their peaks which
influence the strength of the correlations between function points at
different redshifts.
We identify the one which gives the most accurate confidence limits for the number of SNIa we consider. Clearly, for 500 SNIa we can only expect to reconstruct one `feature' in $w$, but it is still important that for more complicated $w$ we trap the function appropriately. 

This work is structured as follows: The basics of GP reconstruction
are summarized in
section~\ref{gaussianprocesses}. Section~\ref{mockdata} lists the
models for which we create mock data sets.
Section~\ref{examples} provides several example reconstructions. In
section~\ref{coverage}, we perform coverage tests which determine the
reliability of the error bars of the reconstructions. These tests
allow us to choose a suitable covariance function. We introduce a
method to determine deviations of a model from the GP reconstructions
in section~\ref{consistency} and conclude in section~\ref{conclusions}.

\section{Gaussian processes}\label{gaussianprocesses}
GP provide a technique to reconstruct a function from observational
data without assuming a specific parameteriziation.  Derivatives of
the function can also be reliably reconstructed. A detailed
description of GP can be found in the book by Rasmussen and Williams
\cite{Rasmussen}, which we will follow throughout this work.  We use
GaPP\footnote{\href{http://www.acgc.uct.ac.za/~seikel/GAPP/index.html}{\url{http://www.acgc.uct.ac.za/~seikel/GAPP/index.html}}}
(Gaussian Processes in Python), which is a freely available GP package developed by
Seikel. It was introduced in \cite{Seikel:2012uu} and has recently
been expanded to allow for marginalization over hyperparameters using
the MCMC package
emcee\footnote{\href{http://github.com/dfm/emcee}{\url{http://github.com/dfm/emcee}}}
\cite{ForemanMackey:2012ig}.

\subsection{Gaussian process reconstructions}\label{gpr}
Assume we have $N$ observational data points $\bm y=(y_1, \dots, y_N)$
at redshifts $\bm Z=(z_1, \dots, z_N)$. The errors of the observations
are given in the covariance matrix $C$. We want to reconstruct the
function underlying the data at points $\bm Z^*=(z^*_1, \dots, z^*_N)$
and denote the function values at these points as $\bm
f^*=(f(z^*_1), \dots, f(z^*_N))$.

GP can be thought of as a generalization of a Gaussian distribution:
At each point $z^*$, the reconstructed function $f(z^*)$ is described
by normal distribution. Function values at different points $z^*$ and
$\tilde{z}^*$ are not independent from each other, but are related by
a covariance function $k(z^*, \tilde{z}^*)$ (see section
\ref{covfunctions} for some choices of $k$). Covariance functions
depend on hyperparameters, such as the signal variance $\sigma_f$ and
the characteristic length scale $\ell$. (More complicated covariance
functions can depend on additional hyperparameters; but in this work,
we will focus on covariance functions that only include $\sigma_f$ and
$\ell$.) In contrast to regular parameters, hyperparameters do not
specify the form of a function; they only characterize typical
changes of the reconstructed function.  $\ell$ roughly corresponds to
the scale in $z$ over which significant changes occur in the function
values and $\sigma_f$ denotes the typical size of these changes. 

For a given covariance function and hyperparameters, the reconstructed
function is determined by the covariances between the data and the
points $\bm Z^*$, where the function is to be reconstructed. The
relevant covariances are given by matrices $K(\bm X,\tilde{\bm X})$
with $\{K(\bm X,\tilde{\bm X})\}_{ij} = k(x_i, \tilde{x}_j)$, where
$\bm X, \tilde{\bm X} \in \{\bm Z, \bm Z^*\}$.

The mean and covariance of the GP reconstruction of the
$n$th derivative of a function are given by
\begin{equation}
\overline{{\bm f}^{*(n)}} = K^{(n,0)}(\bm Z^*,\bm Z) \left[K(\bm Z,\bm Z) +
  C\right]^{-1} {\bm y}
\end{equation}
and
\begin{eqnarray}
&&\hspace{-1em}\text{cov}\left({\bm f}^{*(n)}\right) = K^{(n,n)}(\bm Z^*,\bm Z^*) \\
&&{}- K^{(n,0)}(\bm Z^*,\bm Z) \left[K(\bm Z,\bm Z) +
  C\right]^{-1}  K^{(0,n)}(\bm Z,\bm Z^*) \nonumber
\end{eqnarray}
respectively. Here, $k^{(n,m)}$ denotes the $n$th derivative of $k$
with respect to the first argument and the $m$th derivative with
respect to the second argument.

Besides the data, the above equations also depend on the covariance
function and the values of the hyperparameters. While the covariance
function needs to be chosen by hand, one can either determine the
optimal hyperparameters by maximizing the marginal likelihood
$\mathcal{L} = p({\bm y}|\bm Z,\sigma_f,\ell)$, or one can marginalize
over this likelihood. (The term ``marginal likelihood'' refers to the
fact that one marginalizes over all possible functions $f(z)$.) In
both cases, the hyperparameters are determined by the data. The log
marginal likelihood is given by:
\begin{eqnarray}\label{log-marginal-p}
\ln \mathcal{L} &=& \ln p({\bm y}|\bm Z,\sigma_f,\ell) \nonumber\\
&=& -\frac{1}{2}\bm y^T \left[K(\bm Z,\bm Z) +
  C\right]^{-1} {\bm y} \nonumber \\
&& {} - \frac{1}{2}\ln
  \left|K(\bm Z,\bm Z)+C\right|
  - \frac{n}{2}\ln 2\pi \,. 
\end{eqnarray}
Recall that $K$ depends on the hyperparameters $\sigma_f$ and
$\ell$. 

From a Bayesian point of view, marginalizing over the hyperparameters
is the correct approach. This is, however, computationally much more
expensive than optimizing the hyperparameters. In most cases, the
optimization approach is a good approximation. Problems with this
approximation can occur if the likelihood $\mathcal{L}$ has multiple
(equally high) peaks. However, we have not encountered these problems
when reconstructing dark energy. We show below that the difference between the two is close, and so justify using
the optimization approach.

\subsection{Reconstructing $g(f(z),f'(z),\dots)$}

Often we are not only interested in a function $f(z)$ for which we
have observational data, but also in functions $g(f(z),f'(z),\dots)$
which depend on $f(z)$ and its derivatives. It is important to note
that $f(z)$ and its derivatives are not independent
quantities. Therefore, we need to consider the covariances
\begin{eqnarray}
&&\hspace{-1em}\text{cov}\left({\bm f}^{*(n)}, {\bm f}^{*(m)}\right) = K^{(n,m)}(\bm Z^*,\bm Z^*) \\
&&{}- K^{(n,0)}(\bm Z^*,\bm Z) \left[K(\bm Z,\bm Z) +
  C\right]^{-1}  K^{(0,m)}(\bm Z,\bm Z^*) \nonumber
\end{eqnarray}
when reconstructing $g(f(z),f'(z),\dots)$.

At each point $z^*$, we perform a Monte Carlo sampling to obtain the
distribution of function values for $g^*=g(f^*,f^*{}',\dots)$, where
$f^*=f(z^*)$. In each sampling step, we draw values for
$f^*,f^*{}',\dots$ from a multivariate normal distribution:
\begin{equation}
\begin{bmatrix} {f^*} \\ {f^*{}'} \\ \vdots \end{bmatrix} \sim
\mathcal{N} \left(
\begin{bmatrix}
\overline{f^*}\\
\overline{f^*{}'}\\
\vdots
\end{bmatrix}
,
\begin{bmatrix}
\text{cov}(f^*,f^*)     & \text{cov}(f^*,f^*{}')    & \cdots \\
\text{cov}(f^*,f^*{}')  & \text{cov}(f^*{}',f^*{}') & \cdots \\
\vdots                  & \vdots                   & \ddots
\end{bmatrix}
\right) \,.
\end{equation}
These values are used to calculate $g^*$. Performing many of these
MCMC steps, we obtain a distribution of function values for $g$ at
each redshift $z^*$.

\subsection{Covariance functions}\label{covfunctions}

There is a large variety of covariance functions. The optimal choice
of covariance function depends on the problem at hand (see
\cite{Rasmussen} for some examples). Here, we consider the squared
exponential, which is considered to be a general purpose covariance
function, and the Mat\'ern class, from which individual covariance
functions can be obtained by choosing the parameter $\nu$, which
governs the width of the peak of the covariance function for a given
$\ell$. 

The squared exponential and the Mat\'ern class are given by:
\begin{description}
\item[Squared exponential]
\begin{equation}
k(z,\tilde{z}) =
  \sigma_f^2 \exp\left( - \frac{(z -
    \tilde{z})^2}{2\ell^2} \right) 
\end{equation}
\item[Mat\'ern]
\begin{eqnarray}
k(z,\tilde{z}) &=&
  \sigma_f^2\, \frac{2^{1-\nu}}{\Gamma(\nu)} 
  \left(\frac{\sqrt{2\nu(z - \tilde{z})^2}}{\ell}\right)^\nu
  \nonumber\\
  && {}\times K_\nu \left(\frac{\sqrt{2\nu(z - \tilde{z})^2}}{\ell}\right)
  \,,
\end{eqnarray}
\end{description}
where $K_\nu$ is a modified Bessel function. $\nu$ is a positive
parameter which defines the shape of the covariance
function. While $\nu=1/2$ corresponds to an exponential
$k(z,\tilde{z})=\sigma_f^2 \exp\left( -|z - \tilde{z}|/\ell \right)$,
we recover the squared exponential covariance function for
$\nu\to\infty$.  The Mat\'ern covariance function is $n$-times mean
square differentiable (i.e. the derivative $\partial^{2n}
k(z,\tilde{z})/\partial z^n\partial\tilde{z}^n$ exists and is finite)
if and only if $\nu>n$. A covariance function that is $n$-times mean
square differentiable can be used to reconstuct up to the $n$th
derivative of a function.

For half-integer values of $\nu$, the Mat\'ern covariance function can
be simplified. Here we consider $\nu=5/2$, $\nu=7/2$ and $\nu=9/2$. 
The reconstruction of the equation of state from distance data
requires the reconstruction of the second derivative $D''(z)$ (see
Eq. \eqref{w}). Therefore, the smoothness parameter $\nu$ needs to be
larger than 2.
In the following, the Mat\'ern functions with $\nu=5/2$, $\nu=7/2$ and
$\nu=9/2$ will be refered to as Mat\'ern(5/2), Mat\'ern(7/2) and
Mat\'ern(9/2), respectively. These functions are given by:

\begin{description}
\item[Mat\'ern(5/2)]\begin{eqnarray}
&& k(z,\tilde{z}) = \sigma_f^2
  \exp\left[-\frac{\sqrt{5}\,|z-\tilde{z}|}{\ell}\right] \\
&&\qquad \times \left(1 +
  \frac{\sqrt{5}\,|z-\tilde{z}|}{\ell} +
  \frac{5(z-\tilde{z})^2}{3\ell^2} \right) \nonumber
\end{eqnarray}
\item[Mat\'ern(7/2)]
\begin{eqnarray}
&& k(z,\tilde{z}) = \sigma_f^2
  \exp\left[-\frac{\sqrt{7}\,|z-\tilde{z}|}{\ell}\right] \\
  &&\qquad \times \left(1 +
  \frac{\sqrt{7}\,|z-\tilde{z}|}{\ell} + \frac{14(z-\tilde{z})^2}{5\ell^2} +
  \frac{7\sqrt{7}\,|z-\tilde{z}|^3}{15\ell^3} \right) \nonumber
\end{eqnarray}
\item[Mat\'ern(9/2)]
\begin{eqnarray}
k(z,\tilde{z}) &=& \sigma_f^2
  \exp\left[-\frac{3\,|z-\tilde{z}|}{\ell}\right]  \\
  && \times \left(1 +
  \frac{3\,|z-\tilde{z}|}{\ell} + \frac{27(z-\tilde{z})^2}{7\ell^2}\right. \nonumber\\
&&\qquad \left. {}+ \frac{18\,|z-\tilde{z}|^3}{7\ell^3} +
  \frac{27(z-\tilde{z})^4}{35\ell^4} \right)\nonumber
\end{eqnarray}
\end{description}

\begin{figure}
\includegraphics[width=0.45\textwidth]{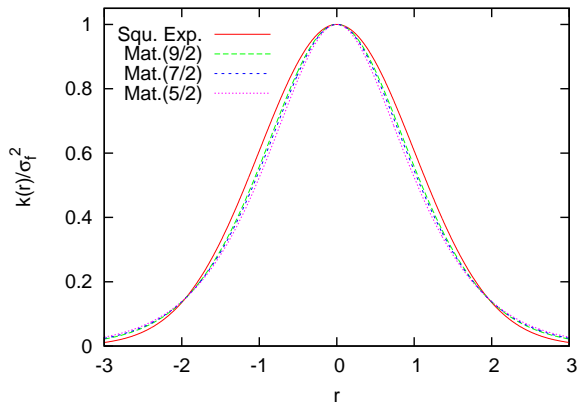}
\caption{Different covariance functions $k(r)/\sigma_f^2$ as a
  function of $r=(z-\tilde{z})/\ell$. Shown are the squared
  exponential and three functions of the Mat\'ern class with $\nu =
  9/2$, $7/2$, $5/2$.}
\label{covfig}
\end{figure}

Figure~\ref{covfig} shows all four covariance functions considered in
this work. Increasing the parameter $\nu$ of the Mat\'ern class widens
the peak of the function. We get the widest peak for the squared
exponential, which corresponds to Mat\'ern($\nu\to\infty$).

\section{Mock data}\label{mockdata}

\begin{figure*}
\subfloat[$w(z)$]{
\includegraphics*[width=0.3\textwidth]{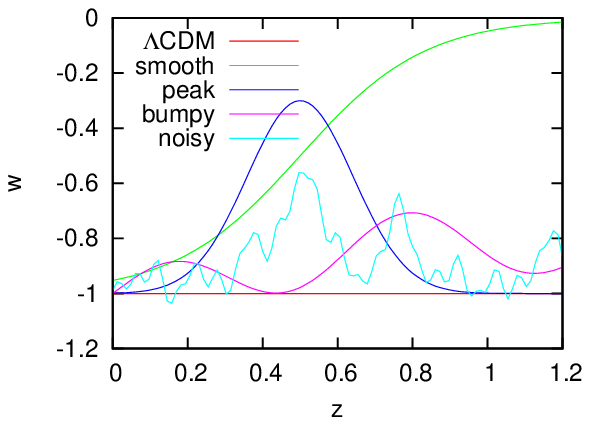}} \hfill
\subfloat[$D(z)$]{
\includegraphics*[width=0.3\textwidth]{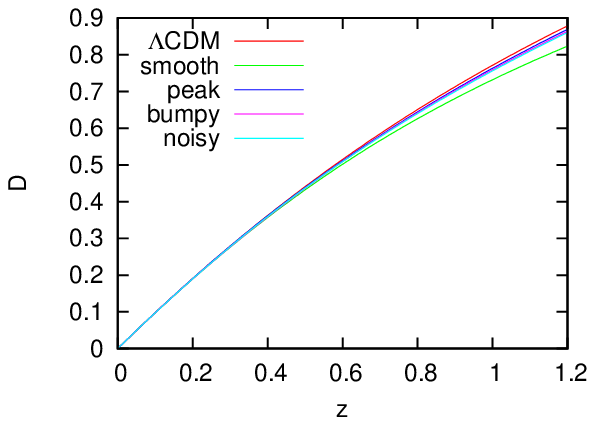}} \hfill
\subfloat[$D(z)-D_{\Lambda\text{CDM}}(z)$]{
\includegraphics*[width=0.3\textwidth]{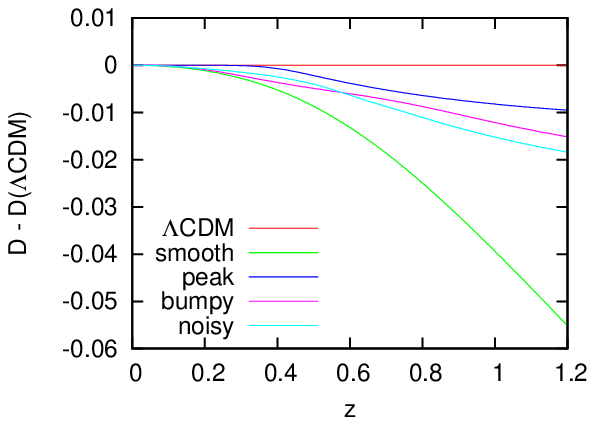}}
\caption{\label{modelsfig}
  $w(z)$, $D(z)$ and $D(z)-D_{\Lambda\text{CDM}}(z)$ for the
  different test models.}
\end{figure*}

We create mock SN data catalogues for different cosmological models.
Each model represents a specific smoothness of the equation of
state, and is not supposed to represent a specific model. $w(z)$ for the five test models is listed in table
\ref{modelstable} and shown in figure \ref{modelsfig}. Also shown are
the dimensionless comoving distance $D(z)$ and the residual distance
$D(z)-D_{\Lambda\text{CDM}}(z)$. For easier reference, we have
assigned the names ``smooth'', ``peak'', ``bumpy'' and ``noisy'' to
the non-$\Lambda$CDM models. We assume all models to be spatially flat
($\Omega_k=0$) and fix the matter density to $\Omega_m=0.3$.

We use the scatter anticipated for the Dark Energy Survey (DES)
\cite{Bernstein:2011zf} to create the mock data. However, instead of
using the anticipated number of $\sim 4000$ SNe, we reduced this
number to 546 to speed up the analysis. For real life observations the
errors are Gaussian in magnitude. Here, we also assume the errors to be
Gaussian in distance $D$ as GP require the observational errors to be
Gaussian. This assumption is justified because the actual errors in $D$
deviate only slightly from Gaussianity for the considered scatter.

\begin{table}
\begin{ruledtabular}
\begin{tabular}{ll}
$\Lambda$CDM & $w(z) = -1$ \\
smooth       & $w(z) = \frac{1}{2} \left(-1 + \tanh\left[3\left(z- \frac{1}{2}\right)\right]\right)$  \\
peak         & $w(z) = -1 + 0.7  \exp\left[-\left(\frac{z-0.5}{0.2}\right)^2\right]$  \\
bumpy        & $w(z) = -0.9 - 0.1 \cos(3.5 z) + 0.1\sin(10 z)$  \\
noisy        & $w(z) = -0.9+0.05 \sin(2.5 z)-0.055 \cos(5.7 z)$ \\
             & \hspace{3em} ${}-0.06 \sin(8.9 z)+0.03 \cos(14.2 z)$ \\
             & \hspace{3em} ${}-0.07 \cos(19.4 z) +0.06 \sin(27.1 z)$ \\
             & \hspace{3em} ${}+0.034 \cos(33.8 z)-0.047 \sin(47.2 z)$ \\
             & \hspace{3em} ${}-0.027 \cos(77.8 z)+0.032 \sin(117.8 z)$
\end{tabular}
\end{ruledtabular}
\caption{\label{modelstable}$w(z)$ for the different test models. From top to bottom the models become increasingly complex, and serve as suitable benchmark models to quantify the sorts of behaviour that can be constrained for a given number of SNIa.}
\end{table}

\section{Example reconstructions}\label{examples}

In this section, we show some example reconstructions. This gives some intuition of how different
methods can affect the result of a reconstruction. However, isolated examples
are not suitable to determine which approach leads to the most
reliable results for a given problem. In order to determine that one
needs to analyze many reconstructions instead of just a few
examples. This is done in sections~\ref{coverage} and
\ref{consistency}.

\subsection{Marginalization vs. optimization}

As mentioned in section~\ref{gpr}, marginalization over the
hyperparameters would be the correct Bayesian approach. As this method
is computationally much more expensive than optimizing the
hyperparameters by maximizing the marginal likelihood
\eqref{log-marginal-p}, it can become prohibitive for large data sets.
Optimization is much faster and in most cases provides good results.
While we focus on the optimization method in this work, we show some
example reconstructions to compare the two different approaches in this
section.

\begin{figure*}
\subfloat[Realisation A \newline Optimization]{
  \includegraphics*[width=0.3\textwidth]{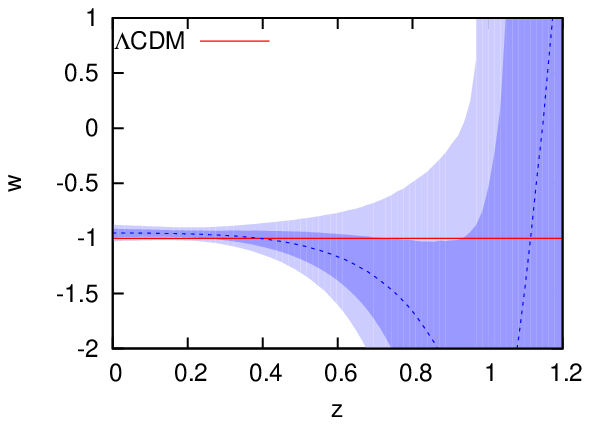}}
\subfloat[Realisation A \newline Marginalization]{
  \includegraphics*[width=0.3\textwidth]{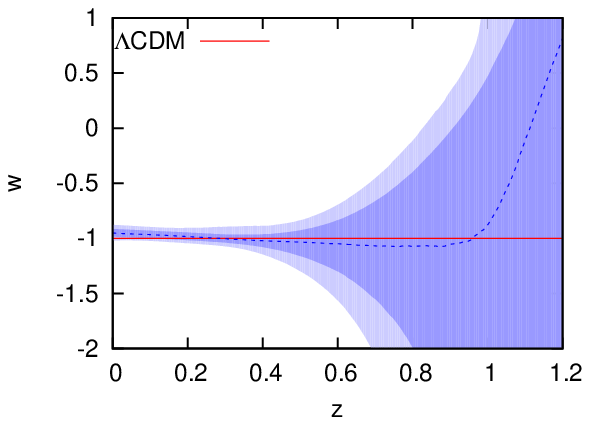}}\\
\subfloat[Realisation B \newline Optimization]{
  \includegraphics*[width=0.3\textwidth]{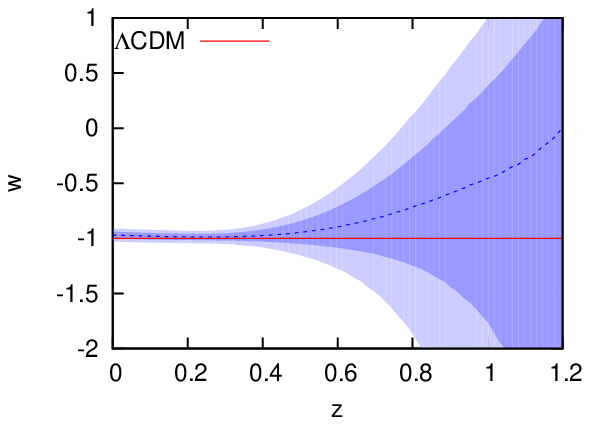}}
\subfloat[Realisation B \newline Marginalization]{
  \includegraphics*[width=0.3\textwidth]{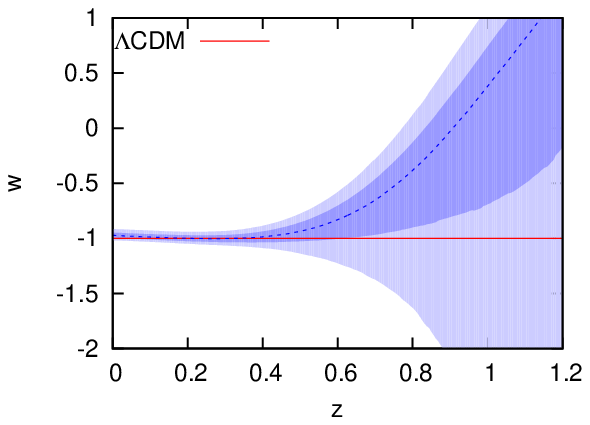}}
\caption{Gaussian process reconstructions of $w(z)$ for two
  realisations of a mock data set. The blue (dotted) line is the mean
  of the reconstruction and the blue shaded regions indicate the 68\%
  and 95\% confidence limits, respectively.  {\em Left:} The
  hyperparameters were optimized for the reconstruction. {\em Right:}
  Marginalization over the hyperparameters using MCMC methods.}
\label{mcmcfig}
\end{figure*}

Figure~\ref{mcmcfig} shows reconstructions of the equation of state
$w(z)$ for two realisations of mock $\Lambda$CDM data sets. Here, we
have used the squared exponential covariance function. The
reconstructions have once been performed by optimizing the
hyperparameters and once by marginalizing over them. In all cases, the
value of the input model, $w=-1$, is captured within the 95\%
confidence limits (CL) of the reconstruction. Note that
marginalization does not necessarily lead to increased errors, which
can be clearly seen for realisation B.

For a given realisation, it is not obvious which approach leads to a
reconstruction that resembles the input model more closely than the
other approach. For realisation A, the reconstructed mean is very
close to the theoretical value up to $z\sim 1$ when marginalizing over
the hyperparameter. Using the optimisation approach, the mean seems
less stable. But for realisation B, the mean and 68\% CL indicate an
increase in $w$ for the marginalisation approach (the 95\% CL still
captures $w=-1$), while optimization resembles $\Lambda$CDM more
closely.

However note that we do expect the input model to leave the 68\% CL
limits for a part of the redshift range for at least some of the
realisations. Therefore, a closer resemblance of the input model does
not necessarily indicate a better method.
See sections~\ref{coverage} and \ref{consistency} for
a detailed analysis.

\subsection{Optimization using different covariance functions}

\begin{figure*}
\subfloat[Squared exponential\newline $\Lambda$CDM]{
  \includegraphics*[width=0.24\textwidth]{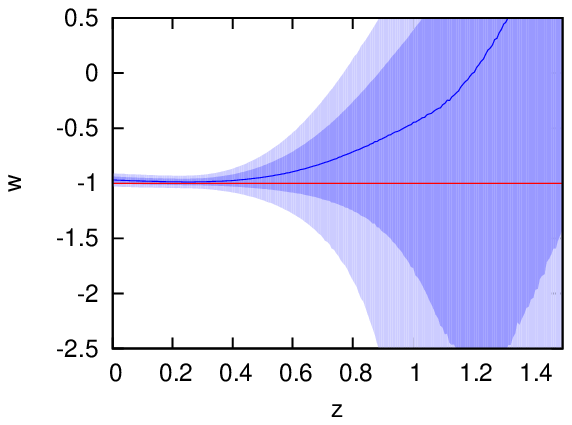}}\hfill
\subfloat[Mat\'ern($9/2$)\newline $\Lambda$CDM]{
  \includegraphics*[width=0.24\textwidth]{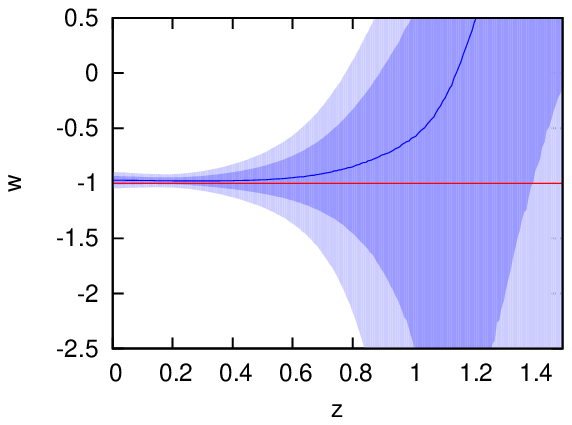}}\hfill
\subfloat[Mat\'ern($7/2$)\newline $\Lambda$CDM]{
  \includegraphics*[width=0.24\textwidth]{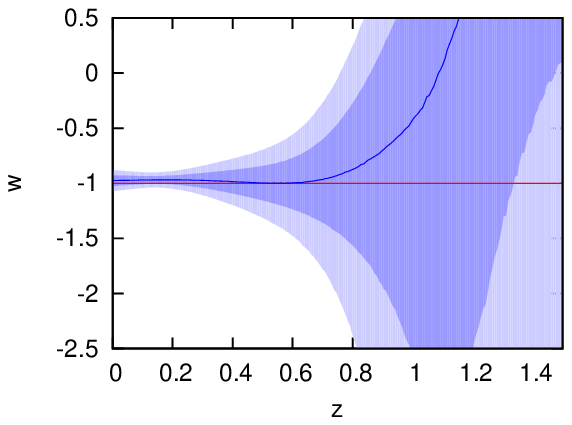}}\hfill
\subfloat[Mat\'ern($5/2$)\newline $\Lambda$CDM]{
  \includegraphics*[width=0.24\textwidth]{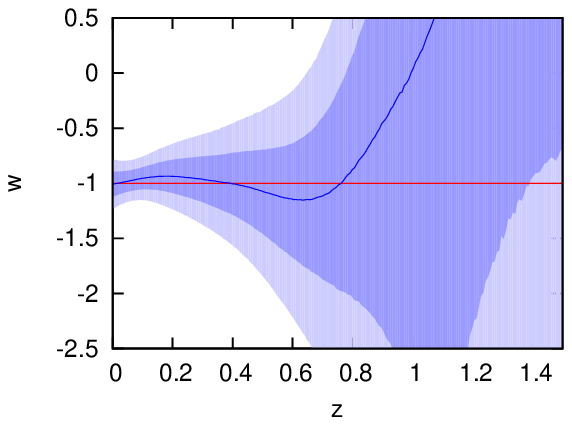}}\\
\subfloat[Squared exponential\newline ``smooth'']{
  \includegraphics*[width=0.24\textwidth]{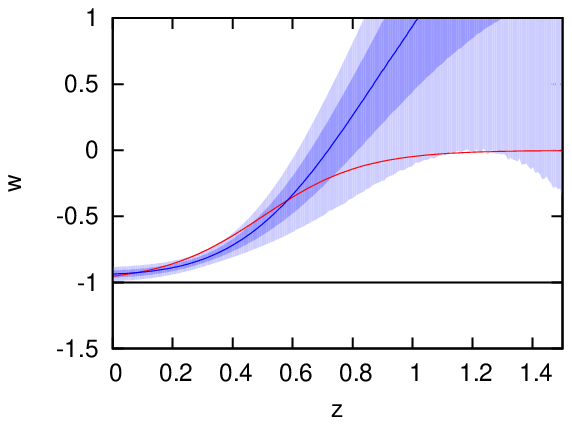}}\hfill
\subfloat[Mat\'ern($9/2$)\newline ``smooth'']{
  \includegraphics*[width=0.24\textwidth]{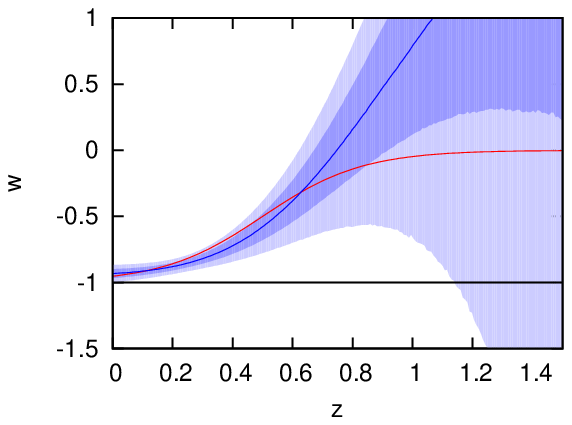}}\hfill
\subfloat[Mat\'ern($7/2$)\newline ``smooth'']{
  \includegraphics*[width=0.24\textwidth]{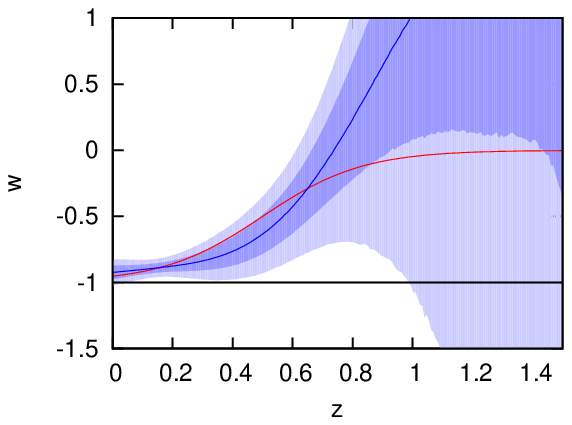}}\hfill
\subfloat[Mat\'ern($5/2$)\newline ``smooth'']{
  \includegraphics*[width=0.24\textwidth]{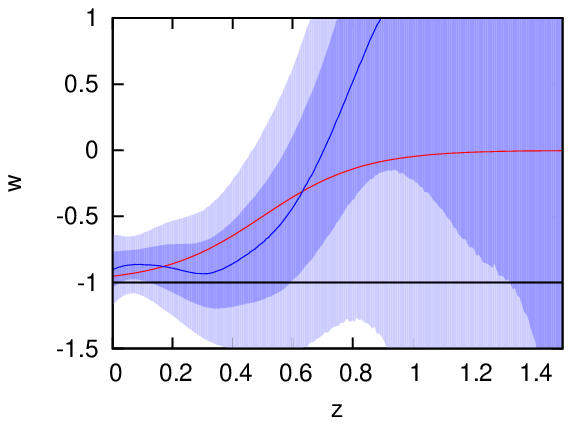}}\\
\subfloat[Squared exponential\newline ``peak'']{
  \includegraphics*[width=0.24\textwidth]{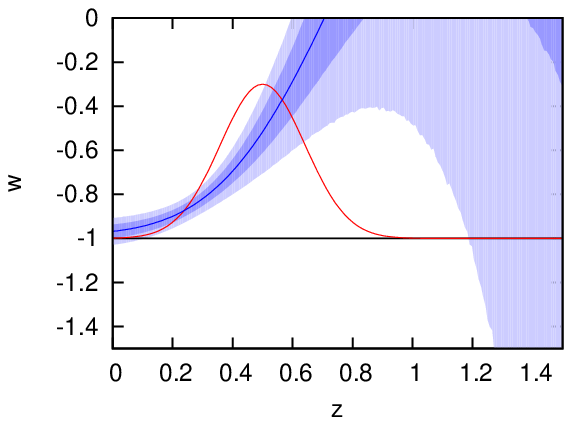}}\hfill
\subfloat[Mat\'ern($9/2$)\newline ``peak'']{
  \includegraphics*[width=0.24\textwidth]{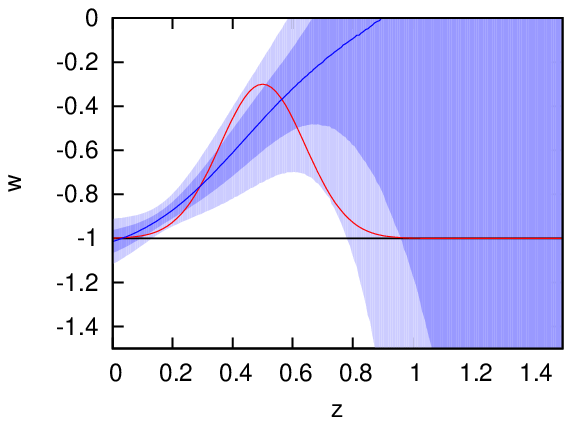}}\hfill
\subfloat[Mat\'ern($7/2$)\newline ``peak'']{
  \includegraphics*[width=0.24\textwidth]{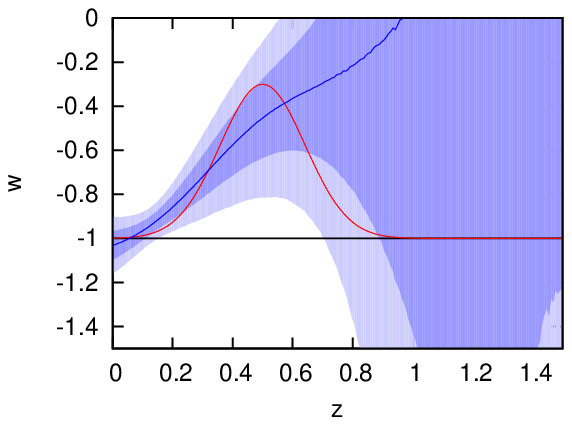}}\hfill
\subfloat[Mat\'ern($5/2$)\newline ``peak'']{
  \includegraphics*[width=0.24\textwidth]{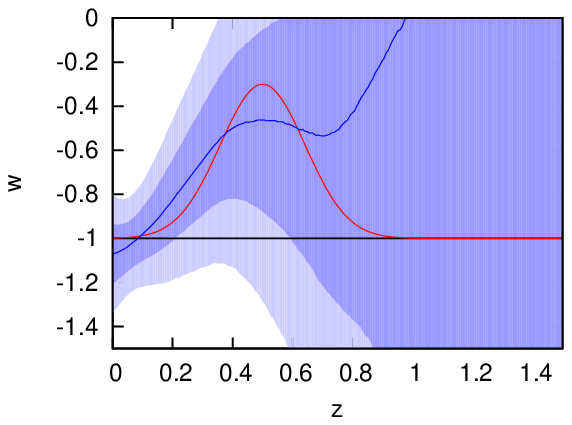}}\\
\subfloat[Squared exponential\newline ``bumpy'']{
  \includegraphics*[width=0.24\textwidth]{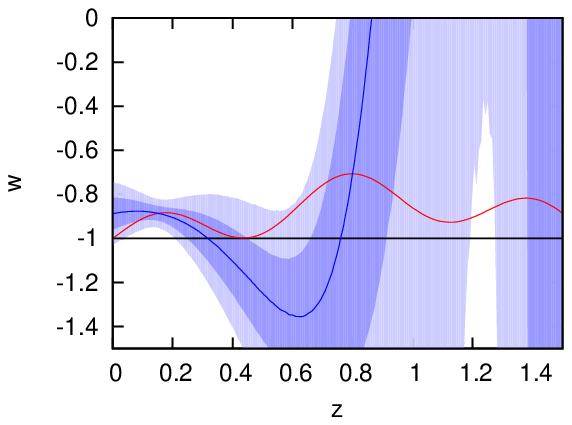}}\hfill
\subfloat[Mat\'ern($9/2$)\newline ``bumpy'']{
  \includegraphics*[width=0.24\textwidth]{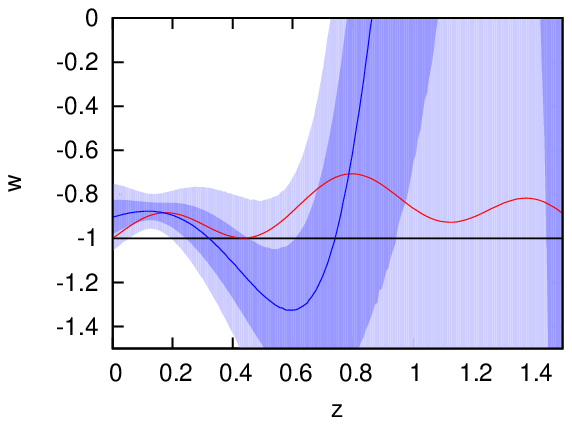}}\hfill
\subfloat[Mat\'ern($7/2$)\newline ``bumpy'']{
  \includegraphics*[width=0.24\textwidth]{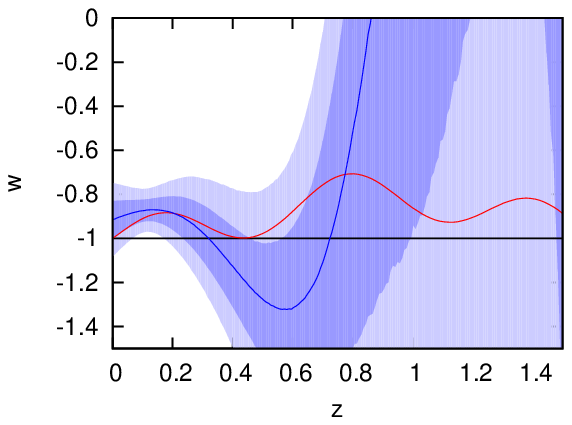}}\hfill
\subfloat[Mat\'ern($5/2$)\newline ``bumpy'']{
  \includegraphics*[width=0.24\textwidth]{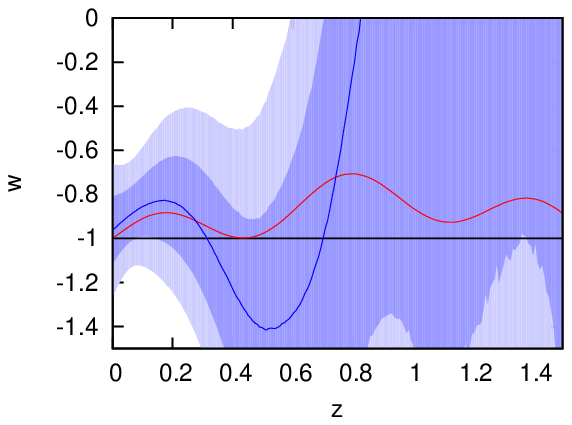}}\\
\subfloat[Squared exponential\newline ``noisy'']{
  \includegraphics*[width=0.24\textwidth]{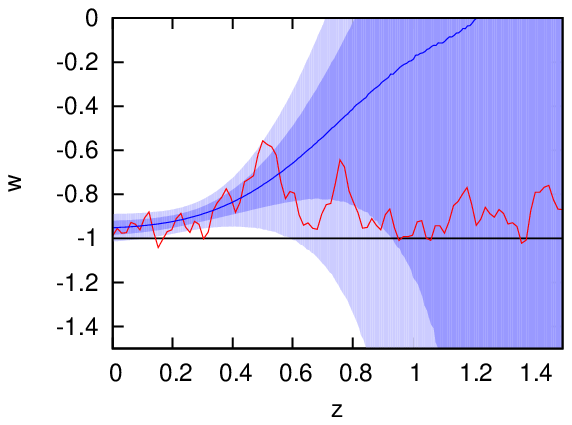}}\hfill
\subfloat[Mat\'ern($9/2$)\newline ``noisy'']{
  \includegraphics*[width=0.24\textwidth]{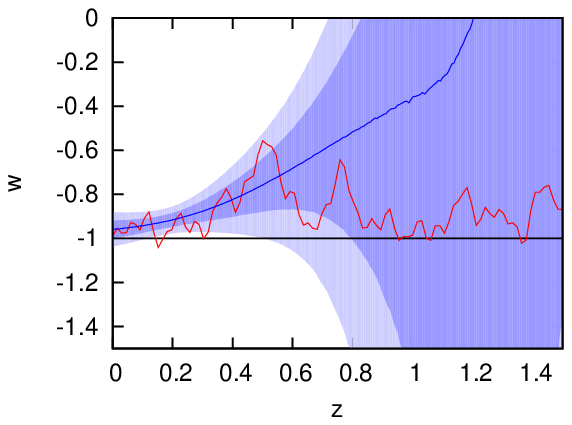}}\hfill
\subfloat[Mat\'ern($7/2$)\newline ``noisy'']{
  \includegraphics*[width=0.24\textwidth]{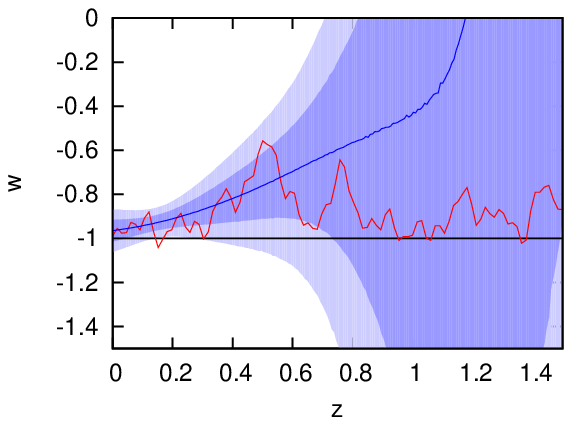}}\hfill
\subfloat[Mat\'ern($5/2$)\newline ``noisy'']{
  \includegraphics*[width=0.24\textwidth]{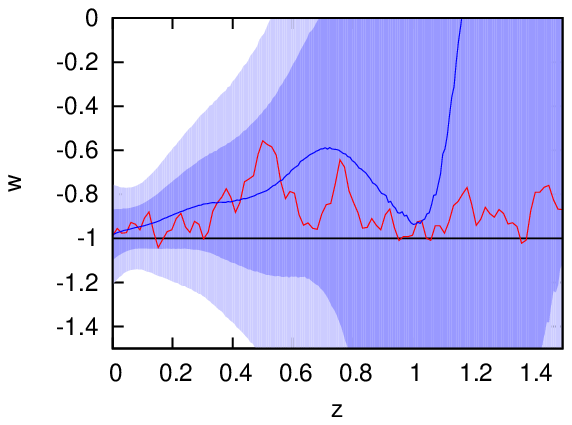}}
\caption{Gaussian process reconstructions of $w(z)$ from a mock data
  set for different cosmological models and using different covariance
  functions. The blue line is the mean of the reconstruction and the
  blue shaded regions indicate the 68\% and 95\% confidence limits,
  respectively. The input model is given by the red line.}
\label{examplesfig}
\end{figure*}

Figure~\ref{examplesfig} shows example reconstructions of $w(z)$ for all
models and covariance functions. The different covariance functions
have been applied to the same realisation of the data for each model.
While the reconstructions using different covariance functions show
similar features for a given realisation, we can also see some
differences. For our examples, these differences are most obvious for
the ``peak'' model. The squared exponential clearly fails to
reconstruct the model, whereas the Mat\'ern functions capture it
within the 95\% CL.

\section{Coverage tests}\label{coverage}

When applying GP to a large variety of problems, the true function is
{\em on average} captured within the 1-$\sigma$ (2-$\sigma$) limits
for 68\% (95\%) of the input range. However, when applying GP to a
specific problem (in our case the reconstruction of dark energy), it
is not guaranteed that we obtain the same amount of coverage. For a
specific problem~-- and in our case any given EOS~-- this coverage depends on the choice of covariance
function. 

\begin{figure*}
\subfloat[Squared exponential\newline $\Lambda$CDM]{
  \includegraphics*[width=0.24\textwidth]{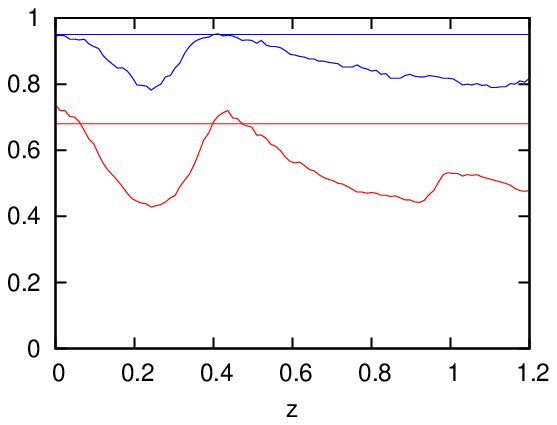}}\hfill
\subfloat[Mat\'ern($9/2$)\newline $\Lambda$CDM]{
  \includegraphics*[width=0.24\textwidth]{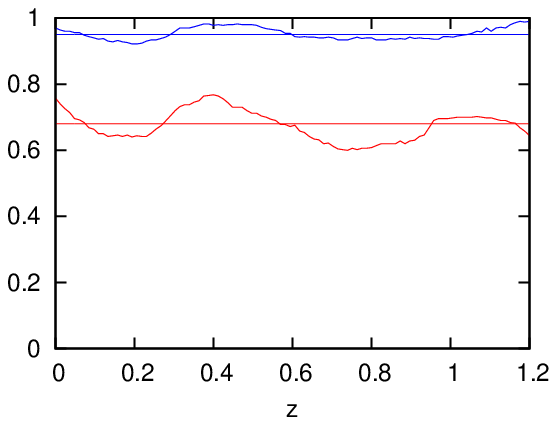}}\hfill
\subfloat[Mat\'ern($7/2$)\newline $\Lambda$CDM]{
  \includegraphics*[width=0.24\textwidth]{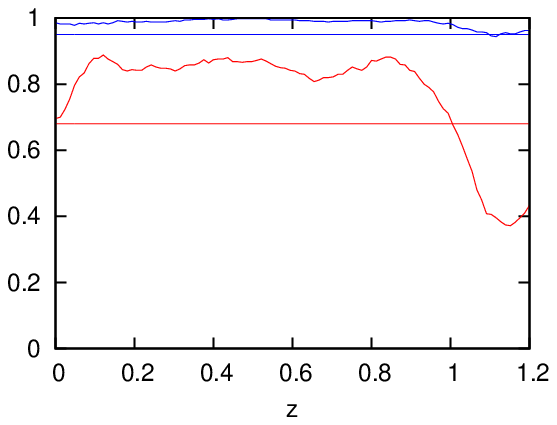}}\hfill
\subfloat[Mat\'ern($5/2$)\newline $\Lambda$CDM]{
  \includegraphics*[width=0.24\textwidth]{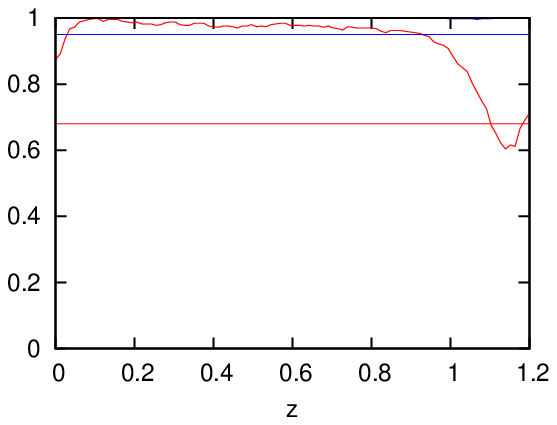}}
\\
\subfloat[Squared exponential\newline ``smooth'']{
  \includegraphics*[width=0.24\textwidth]{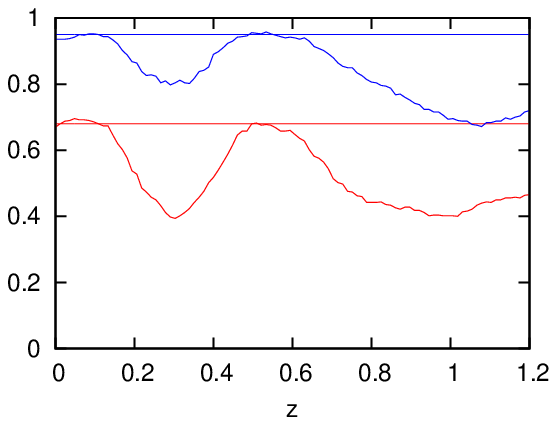}}\hfill
\subfloat[Mat\'ern($9/2$)\newline ``smooth'']{
  \includegraphics*[width=0.24\textwidth]{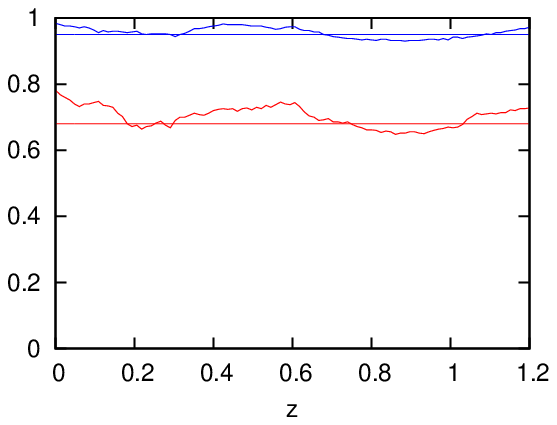}}\hfill
\subfloat[Mat\'ern($7/2$)\newline ``smooth'']{
  \includegraphics*[width=0.24\textwidth]{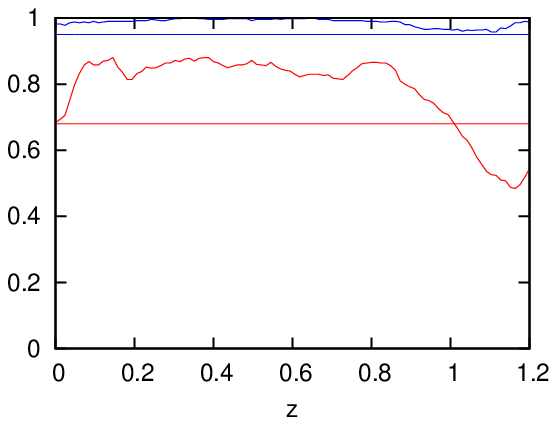}}\hfill
\subfloat[Mat\'ern($5/2$)\newline ``smooth'']{
  \includegraphics*[width=0.24\textwidth]{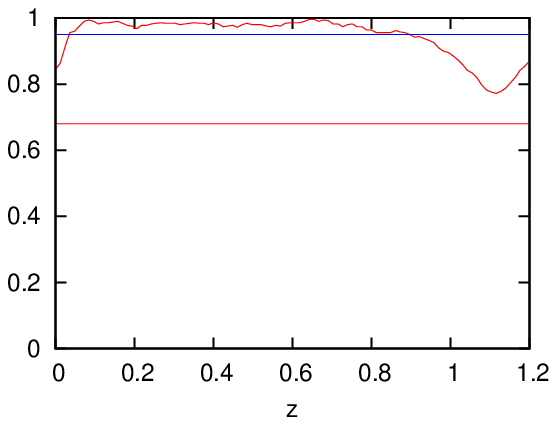}}
\\
\subfloat[Squared exponential\newline ``peak'']{
  \includegraphics*[width=0.24\textwidth]{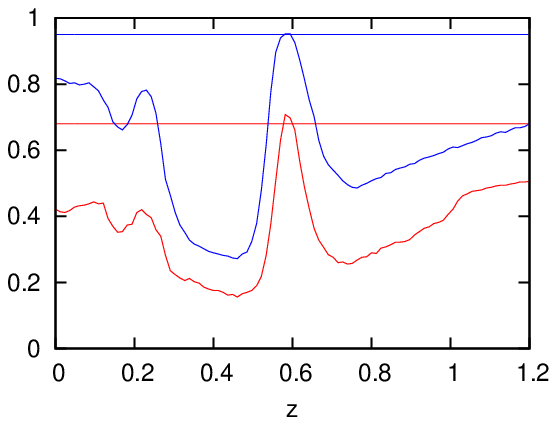}}\hfill
\subfloat[Mat\'ern($9/2$)\newline ``peak'']{
  \includegraphics*[width=0.24\textwidth]{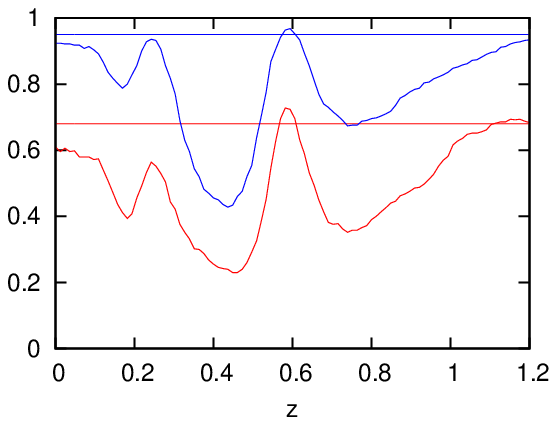}}\hfill
\subfloat[Mat\'ern($7/2$)\newline ``peak'']{
  \includegraphics*[width=0.24\textwidth]{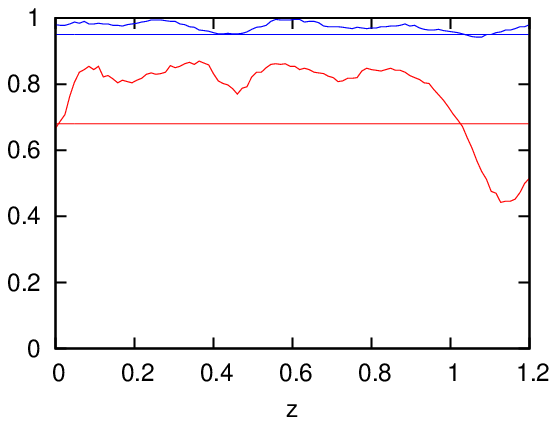}}\hfill
\subfloat[Mat\'ern($5/2$)\newline ``peak'']{
  \includegraphics*[width=0.24\textwidth]{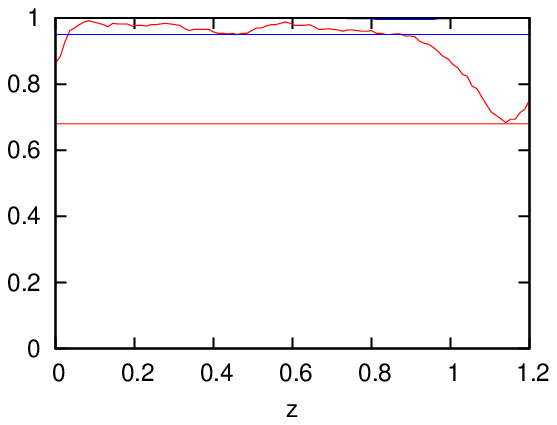}}
\\
\subfloat[Squared exponential\newline ``bumpy'']{
  \includegraphics*[width=0.24\textwidth]{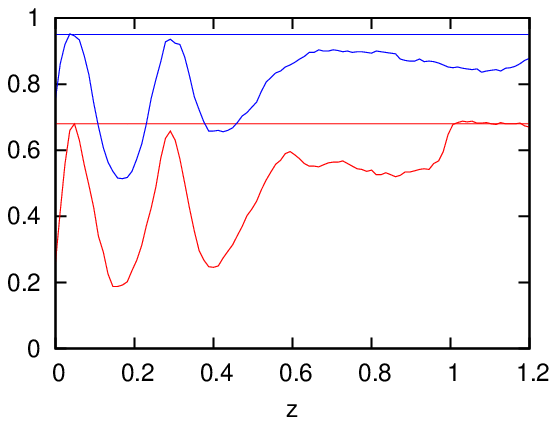}}\hfill
\subfloat[Mat\'ern($9/2$)\newline ``bumpy'']{
  \includegraphics*[width=0.24\textwidth]{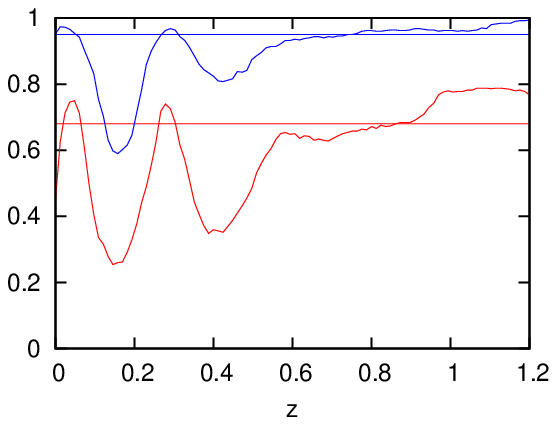}}\hfill
\subfloat[Mat\'ern($7/2$)\newline ``bumpy'']{
  \includegraphics*[width=0.24\textwidth]{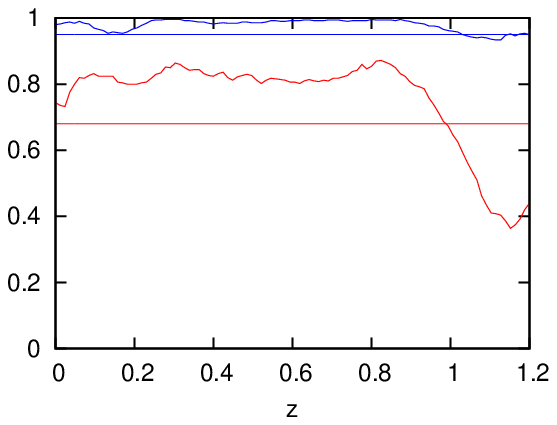}}\hfill
\subfloat[Mat\'ern($5/2$)\newline ``bumpy'']{
  \includegraphics*[width=0.24\textwidth]{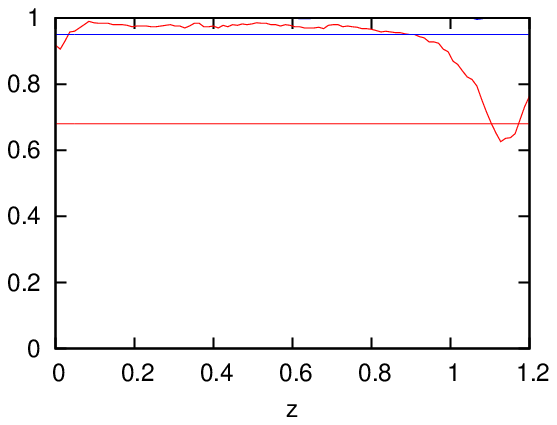}}
\\
\subfloat[Squared exponential\newline ``noisy'']{
  \includegraphics*[width=0.24\textwidth]{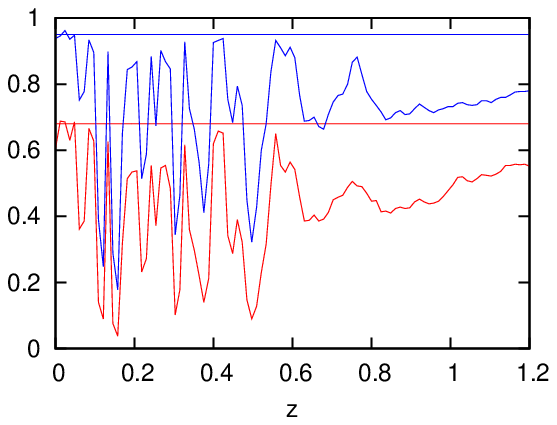}}\hfill
\subfloat[Mat\'ern($9/2$)\newline ``noisy'']{
  \includegraphics*[width=0.24\textwidth]{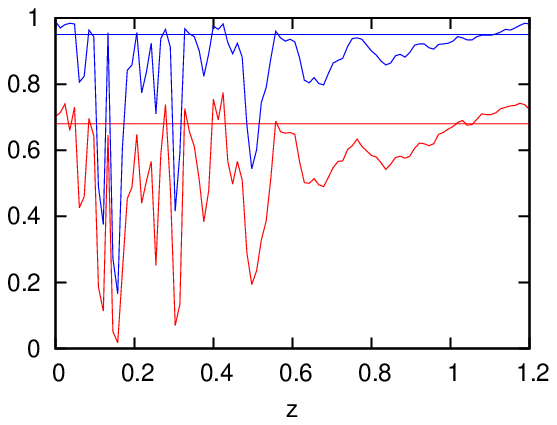}}\hfill
\subfloat[Mat\'ern($7/2$)\newline ``noisy'']{
  \includegraphics*[width=0.24\textwidth]{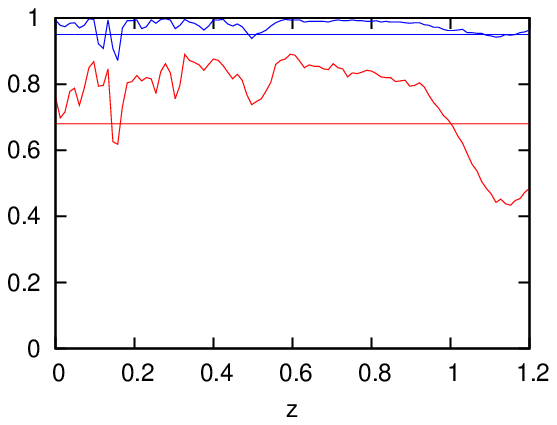}}\hfill
\subfloat[Mat\'ern($5/2$)\newline ``noisy'']{
  \includegraphics*[width=0.24\textwidth]{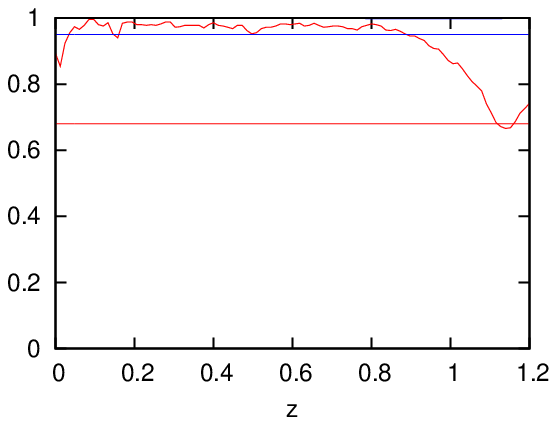}}
\caption{Coverage test for $w(z)$ using 500 mock data sets for
  different cosmological models and using different covariance
  functions. Shown is the fraction of input model points captured with
  the 1-$\sigma$ (red) and 2-$\sigma$ (blue) limits of the
  reconstructions. The constant red and blue lines indicate the
  theoretically expected values 0.68 and 0.95, respectively.}
\label{coveragefig}
\end{figure*}

For each model, we created 500 mock data sets and performed
reconstructions of $w(z)$ for all data sets and covariance
functions. Averaging over all realisations for each model, we then
determined the fraction of input function points lying between the
reconstructed 1- and 2-$\sigma$ limits, respectively, for each
covariance function and each redshift point. The result is shown in
figure~\ref{coveragefig}. 

Table~\ref{coveragetab} lists these fractions averaged over redshift
and also shows the results for the reconstructions of $D$, $D'$, $D''$
and $q$. The covariance functions are listed in the order of
decreasing peak width, i.e.\ decreasing $\nu$. Recall that the squared
exponential corresponds to Mat\'ern($\infty$). GP are a distribution
over functions, which consists of functions of different
smoothness. Covariance functions with wider peak lead to a stronger
preference of smooth functions in this distribution due to the
stronger correlation of function points for $|z-\tilde{z}|\lesssim
2\ell$ (see figure~\ref{covfig}). Scales larger than $2\ell$ are not
relevant for our purposes as the considered redshift range is much
smaller than $2\ell$ (using the optimized value of $\ell$) in all
cases.

The preference of smooth functions in the GP distribution subsequently
leads to smaller error bars for the reconstruction. Thus, for a given
input model, we expect lower coverage for covariance functions with
wider peaks due to the smaller error bars of the reconstruction. This
effect can indeed be seen in the results of table~\ref{coveragetab}.

For $\Lambda$CDM and the ``smooth'' model, Mat\'ern($9/2$) leads to
the most realistic error bars in the sense that the coverage obtained
by the simulation is very close to the theoretically expected values
of 68\% and 95\% for the 1- and 2-$\sigma$ limits, respectively. When
using the squared exponential, we obtain a smaller coverage,
indicating that the errors are underestimated. In contrast, the errors
are overestimated for Mat\'ern($7/2$) and Mat\'ern($5/2$).

For the models with stronger redshift dependence (``peak'', ``bumpy''
and ``noisy''), Mat\'ern($9/2$) and Mat\'ern($7/2$) come closest to
the theoretically expected values, with Mat\'ern($9/2$)
underestimating and Mat\'ern($7/2$) overestimating the error bars.

\begin{table*}
\begin{ruledtabular}
\begin{tabular}{lllll | llll}
       & Squared exp. & Mat\'ern($9/2$) & Mat\'ern($7/2$) &
Mat\'ern($5/2$) & Squared exp. & Mat\'ern($9/2$) & Mat\'ern($7/2$) &
Mat\'ern($5/2$) \\
\hline
$\Lambda$CDM &&&& &&&& \\
$D$    & 0.636   & {\bf 0.676} & 0.685   & 0.728  & 0.924    & {\bf 0.950} & 0.952   & 0.968 \\
$D'$   & 0.580   & {\bf 0.661} & 0.712   & 0.793  & 0.899    & {\bf 0.947} & 0.963   & 0.985 \\
$D''$  & 0.518   & {\bf 0.684} & 0.770   & 0.934  & 0.851    & {\bf 0.960} & 0.989   & 1.000 \\
$w$    & 0.542   & {\bf 0.676} & 0.776   & 0.929  & 0.863    & {\bf 0.953} & 0.985   & 1.000 \\
$q$    & 0.521   & {\bf 0.678} & 0.774   & 0.937  & 0.850    & {\bf 0.954} & 0.987   & 1.000  \\
\hline
Smooth &&&& &&&& \\
$D$    & 0.663   & 0.703       & {\bf 0.696} & 0.723  & 0.935  & 0.955  & {\bf 0.952} & 0.966  \\
$D'$   & 0.595   & {\bf 0.695} & 0.727       & 0.794  & 0.904  & {\bf 0.953}  & 0.966  & 0.985  \\
$D''$  & 0.505   & {\bf 0.707} & 0.784       & 0.940  & 0.823  & {\bf 0.963}  & 0.988  & 1.000  \\
$w$    & 0.524   & {\bf 0.702} & 0.785       & 0.945  & 0.836  & {\bf 0.956}  & 0.987  & 1.000  \\
$q$    & 0.516   & {\bf 0.702} & 0.787       & 0.945  & 0.833  & {\bf 0.957}  & 0.987  & 1.000  \\
\hline
Peak &&&& &&&&\\
$D$    & 0.544   & 0.604   & {\bf 0.691}   & 0.724  & 0.868  & 0.912   & {\bf 0.950} & 0.965  \\
$D'$   & 0.437   & 0.539   & {\bf 0.720}   & 0.788  & 0.739  & 0.850   & {\bf 0.961} & 0.980  \\
$D''$  & 0.337   & 0.480   & {\bf 0.770}   & 0.927  & 0.568  & 0.773   & {\bf 0.973} & 1.000  \\
$w$    & 0.361   & 0.491   & {\bf 0.774}   & 0.926  & 0.598  & 0.781   & {\bf 0.973} & 0.999  \\
$q$    & 0.344   & 0.482   & {\bf 0.775}   & 0.932  & 0.575  & 0.770   & {\bf 0.973} & 0.999  \\
\hline
Bumpy &&&& &&&&\\
$D$    & 0.621  & 0.663  & {\bf 0.679}   & 0.717  & 0.925  &  0.931   & {\bf 0.950} & 0.968 \\
$D'$   & 0.569  & 0.638  & {\bf 0.707}   & 0.783  & 0.890  &  0.929   & {\bf 0.961} & 0.983  \\
$D''$  & 0.479  & 0.605  & {\bf 0.751}   & 0.931  & 0.793  &  0.902   & {\bf 0.981} & 1.000  \\
$w$    & 0.506  & {\bf 0.605}  & 0.756   & 0.929  & 0.812  &  0.906   & {\bf 0.978} & 1.000  \\
$q$    & 0.482  & 0.601  & {\bf 0.755}   & 0.936  & 0.795  &  0.899   & {\bf 0.979} & 1.000  \\
\hline
Noisy &&&& &&&&\\
$D$    & 0.621  & {\bf 0.671}  & 0.693  & 0.718  & 0.927  & {\bf 0.953} & 0.956   & 0.964  \\
$D'$   & 0.565  & {\bf 0.649}  & 0.720  & 0.793  & 0.875  & {\bf 0.941} & 0.964   & 0.981  \\
$D''$  & 0.406  & 0.544  & {\bf 0.752}  & 0.934  & 0.698  & 0.853   & {\bf 0.977} & 1.000  \\
$w$    & 0.440  & 0.559  & {\bf 0.760}  & 0.929  & 0.733  & 0.865   & {\bf 0.975} & 0.999  \\
$q$    & 0.414  & 0.547  & {\bf 0.757}  & 0.937  & 0.704  & 0.852   & {\bf 0.975} & 1.000  \\
\end{tabular}
\end{ruledtabular}
\caption{Coverage test of the Gaussian process for different models
  and covariance functions. {\em Left:} Fraction of function points
  lying between the 1-$\sigma$ limits. In each line, the fraction
  closest to the theoretical value 0.68 has been highlighted. 
  {\em Right:} Fraction of function points lying between the
  2-$\sigma$ limits. In each line, the fraction closest to the
  theoretical value 0.95 has been highlighted.}
\label{coveragetab}
\end{table*}

\section{Consistency with a model}\label{consistency}

GP provide a method for non-parametric reconstructions directly from
observational data. But often one would like to know whether a
specific model is consistent with the GP reconstruction. In this
section, we will describe how deviations of a model from the GP
reconstructions can be quantified. We explain the method by means of
an example, namely $\Lambda$CDM. The technique can be applied to other
models analogously.  For the reconstructions, we use the
Mat\'ern($9/2$) covariance function, which was shown in
section~\ref{coverage} to provide the most reliable results for our
purposes.

We create 1000 mock $\Lambda$CDM data sets, using the scatter and
redshift distributions as described in section~\ref{mockdata}. For
each data set, we reconstruct $w$ and determine the fraction of the
redshift range, in which the fiducial $\Lambda$CDM model is captured
by the 1-, 2- and 3-$\sigma$ limits, corresponding to 68.3\%, 95.4\% and
99.7\% CL, respectively of the reconstruction. Note that here
we determine this fraction for each realisation of the data
individually, which is in contrast to the averaging over all mock data
sets that has been done in Section~\ref{coverage}.

\begin{figure*}
\subfloat[1-$\sigma$]{
  \includegraphics*[width=0.32\textwidth]{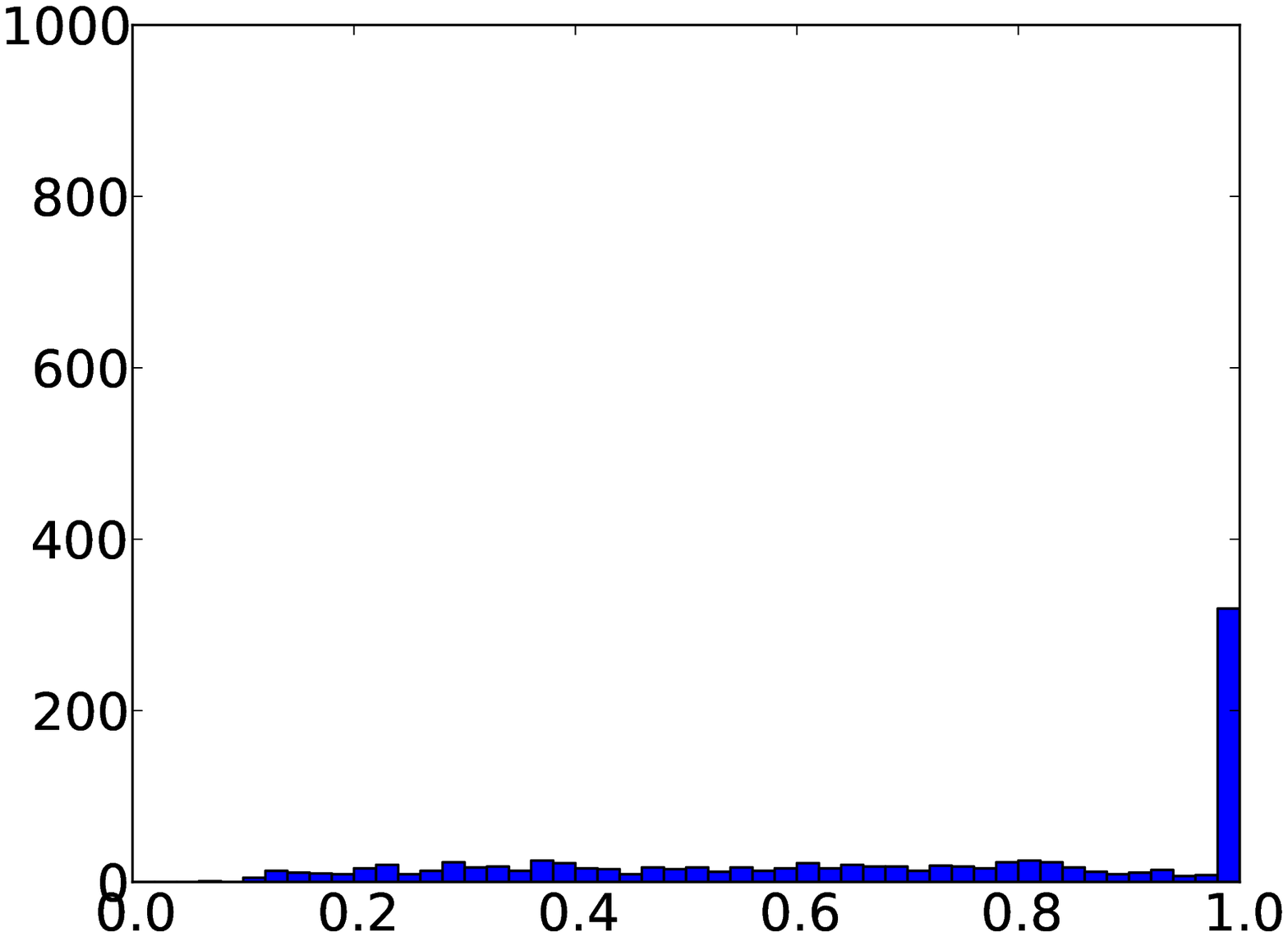}} \hfill
\subfloat[2-$\sigma$]{
  \includegraphics*[width=0.32\textwidth]{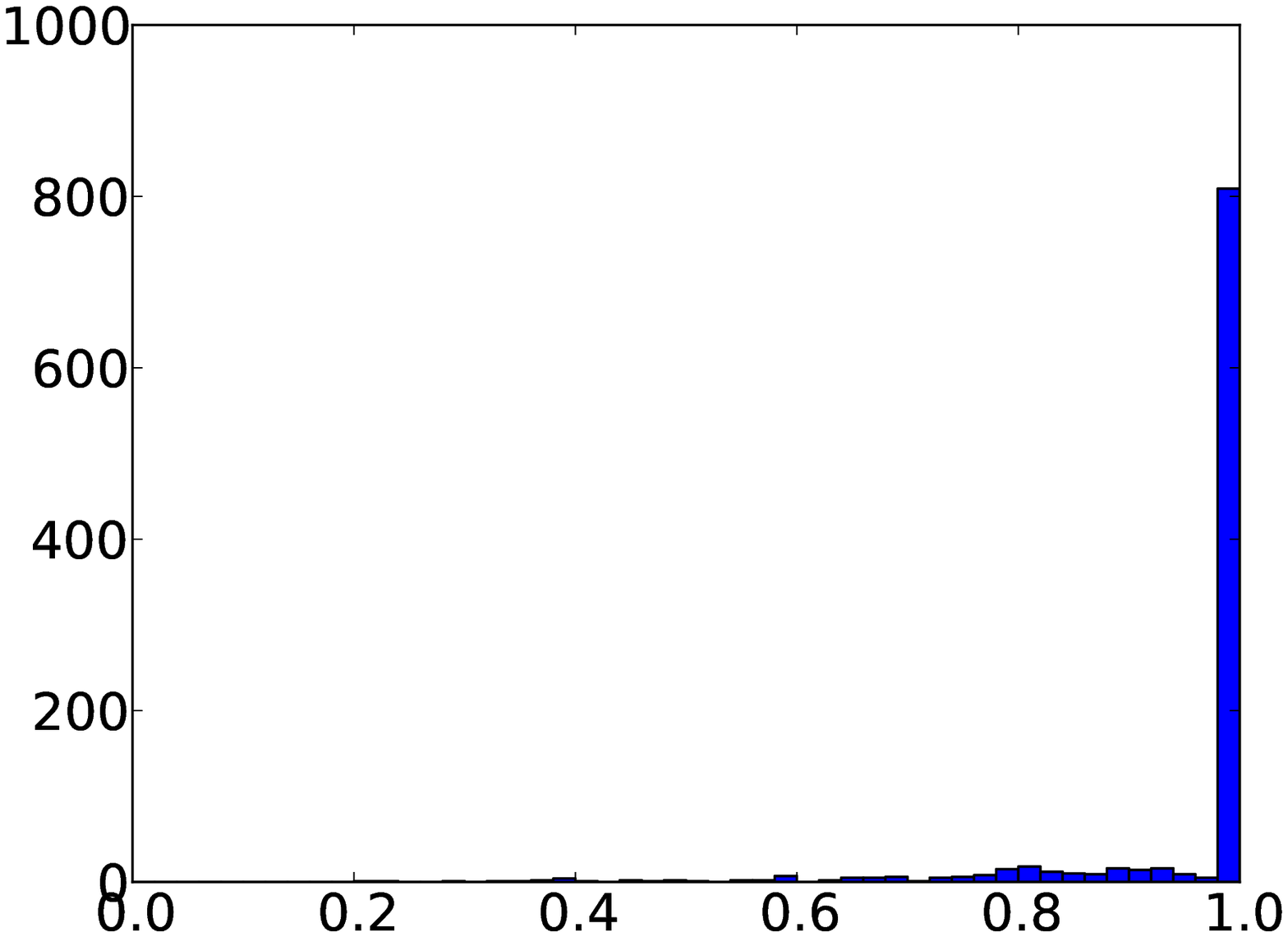}} \hfill
\subfloat[3-$\sigma$]{
  \includegraphics*[width=0.32\textwidth]{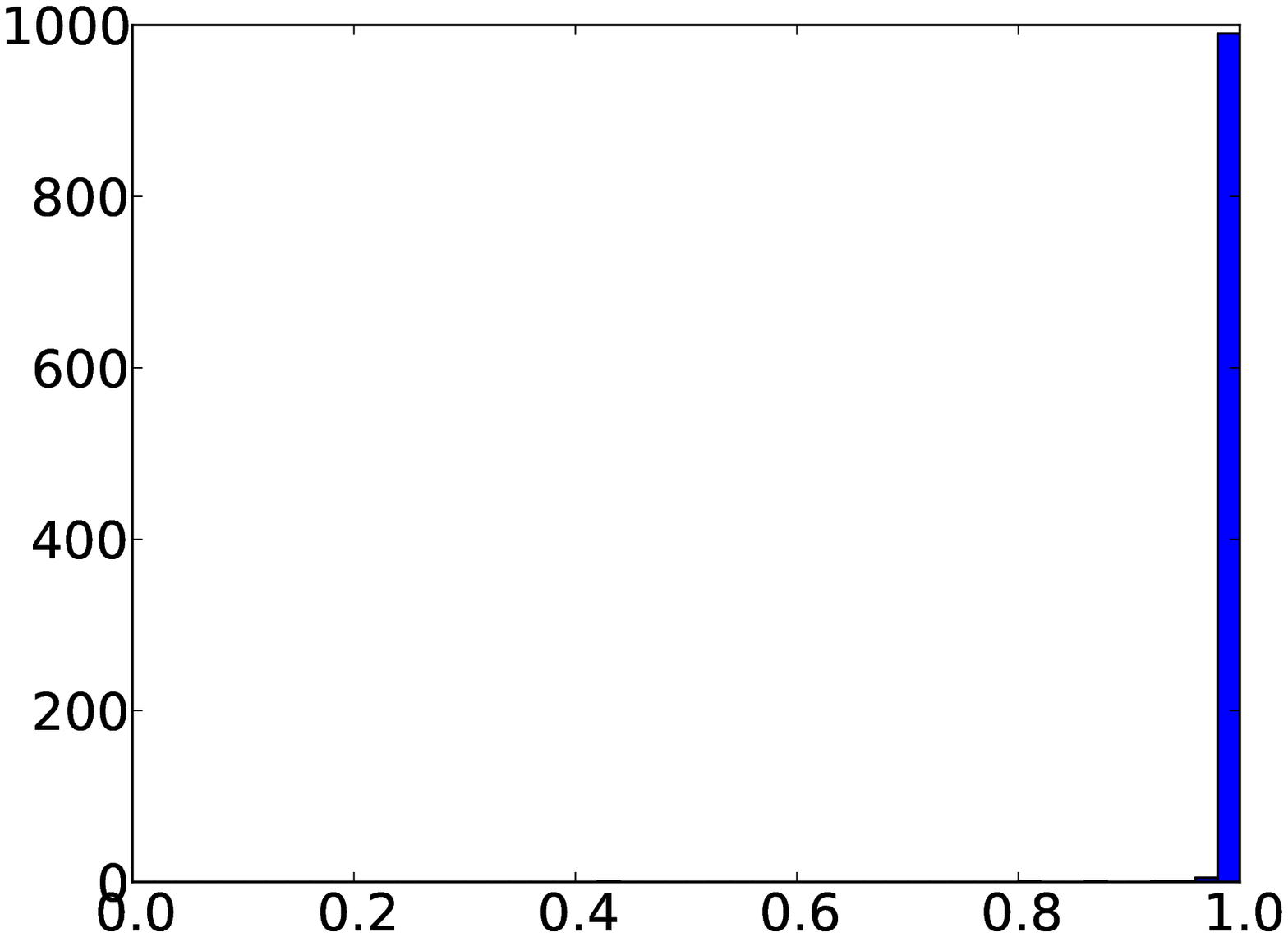}}\\
\subfloat[1-$\sigma$]{
  \includegraphics*[width=0.32\textwidth]{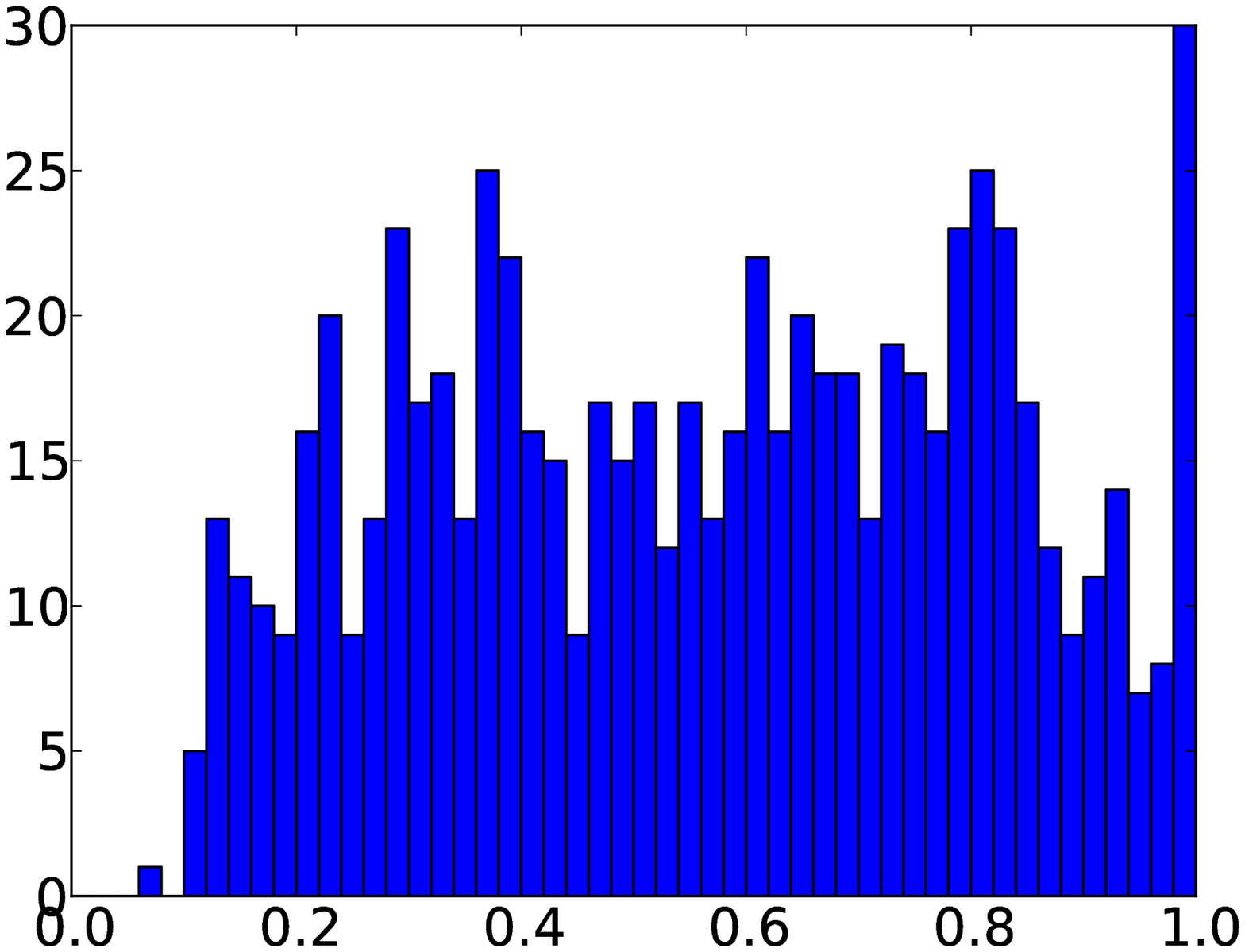}} \hfill
\subfloat[2-$\sigma$]{
  \includegraphics*[width=0.32\textwidth]{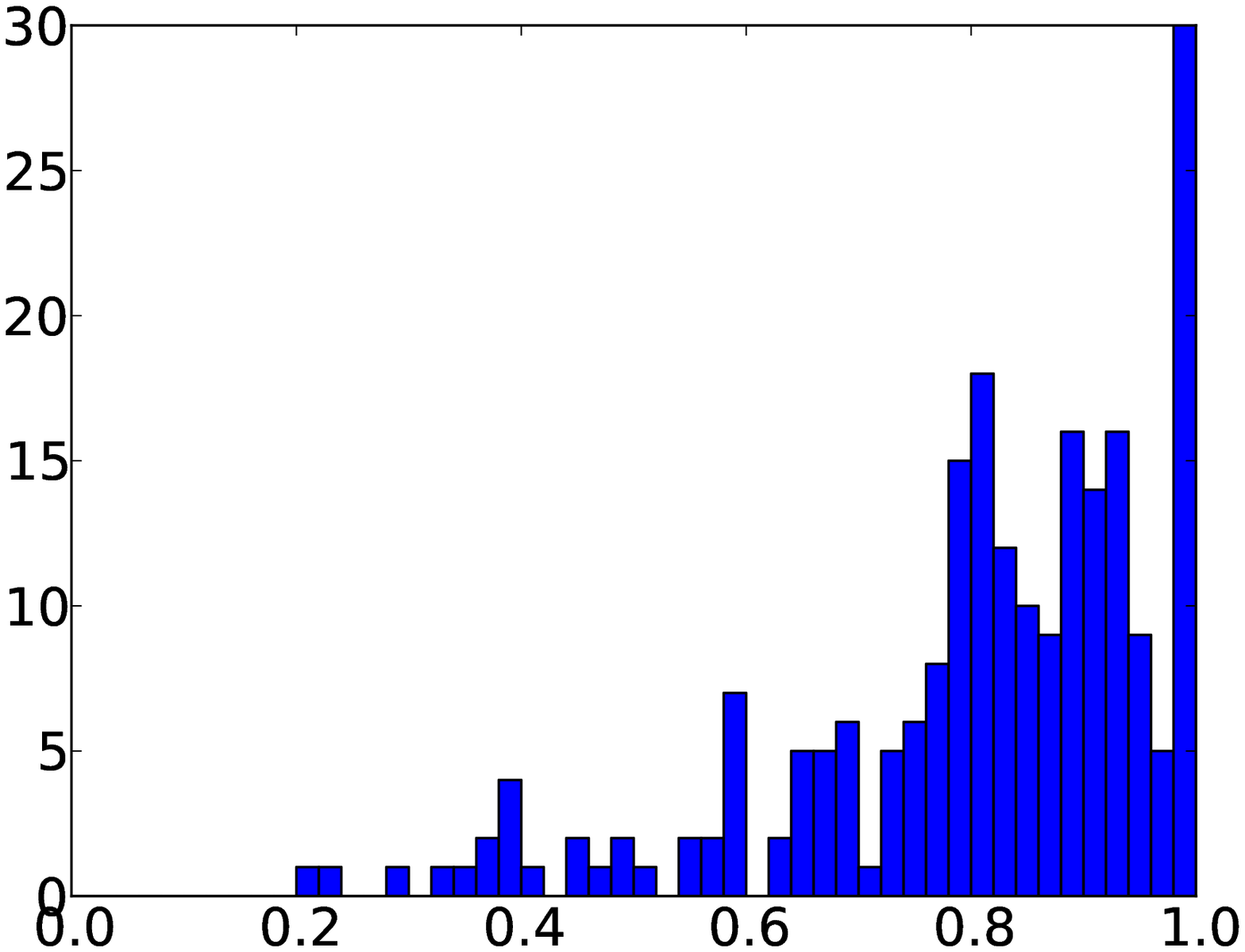}} \hfill
\subfloat[3-$\sigma$]{
  \includegraphics*[width=0.32\textwidth]{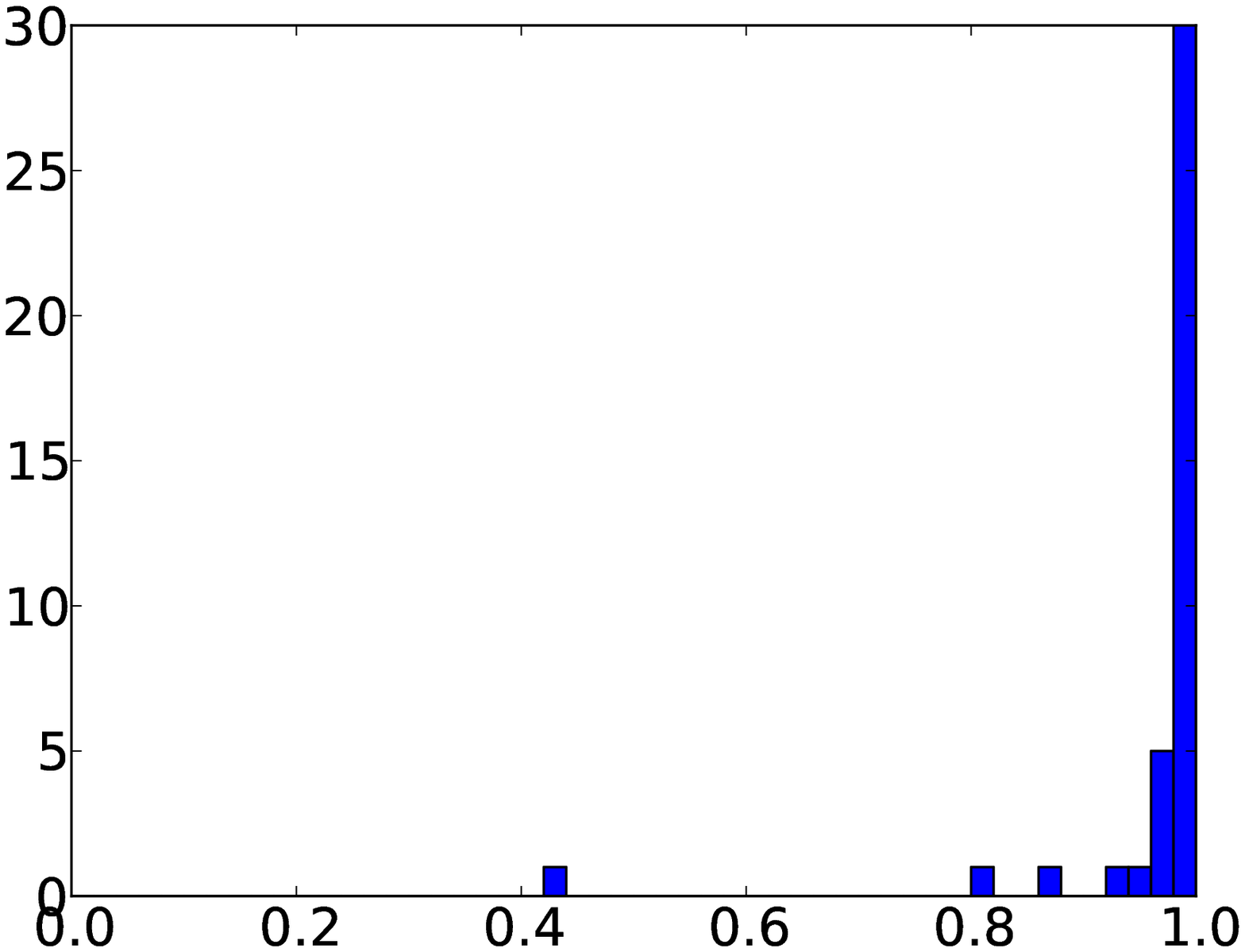}}
\caption{Histograms of the percentages of the redshift range, where
  the fiducial $\Lambda$CDM model is captured by the 1-, 2- and
  3-$\sigma$ limits of the reconstructions of $w$ for a given
  realisation of the data. In total, 1000 mock data sets have been
  used. The bottom row shows a zoomed-in version of the plots in the
  top row.}
\label{coveragehist}
\end{figure*}

Figure~\ref{coveragehist} summarizes the obtained fractions as
histograms.  For roughly one third of the mock data sets, the fiducial
model is captured within the 1-$\sigma$ limits of the reconstruction
over the complete redshift range. For 983 out of 1000 data sets, it is
captured within the 3-$\sigma$ limits over the whole redshift
range. While we expect the fiducial model to lie within the 1-$\sigma$
limits {\em on average} over 68\% of the redshift range, the histgram
shows a very large spread of the actual percentages for the individual
realisations of the data. For a given data set, it is not unusual to
find that the fiducial model is captured within the 1-$\sigma$ limits
anywhere between 10\% and 100\% of the redshift range. When we
consider the 2-$\sigma$ limits, small percentages become less
likely. For the 3-$\sigma$ limits, we can expect a coverage close to
100\%.

We can use these histograms to quantify deviations of the
reconstructions from $\Lambda$CDM when applying GP to a data set,
which may or may not represent $\Lambda$CDM. As the 1-$\sigma$ limits
allow for a large range of percentages and thus offer very little
constraints, the 2- and 3-$\sigma$ limits are more suitable for this
test. For example, if we find that $\Lambda$CDM is captured within the
3-$\sigma$ limits of the reconstruction over 95\% of the redshift
range, we can exclude this model at a 99.6\% CL as we only found
smaller coverage values for 4 out of 1000 mock data sets. Or if
$\Lambda$CDM is captured within the 2-$\sigma$ limits over only 50\% of
the redshift range, then $\Lambda$CDM can be ruled out at a 98.3\% CL.

In order to quantify deviations for a different model, we would need
to produce histograms using mock data sets which assume exactly that
model. Also note that when applying the method to a real world
problem, it is necessary to create mock data sets using the scatter
and redshifts of the actual data set.

\section{Conclusions}\label{conclusions}

We analyzed various aspects of using GP to reconstruct dark energy
from distance measurements. The main questions addressed in this paper
were how to choose a suitable covariance function and how to quantify
deviations of a model from the GP reconstruction. The presented
methods can be applied to other problems analogously.

In theory, the fiducial model that was used to create a mock data set
should be captured within the 1-$\sigma$ (2-$\sigma$) limits of the GP
reconstruction for 68\% (95\%) of the input range. These are average
values when GP are applied to a variety of different problems. These
values do not necessarily apply if we use GP for a specific problem ---
in our case the reconstruction of dark energy from SN data. We tested
various covariance functions on mock data sets for different
cosmological models to determine which function recovers the
theoretical coverage values for the reconstruction of $w$. 

While Mat\'ern($7/2$) provides very good results for the
reconstruction of $D$ for all models, it overestimates the errors for
the reconstructions of the derivatives of $D$. If one is only
interested in $D$ itself, but not in its derivatives or any derived
quantities, then Mat\'ern($7/2$) is the best choice.  

We found that the Mat\'ern($9/2$) covariance function leads to
excellent results for $\Lambda$CDM and the ``smooth'' model for all
reconstructed quantities, but tends to underestimate the errors for
the other models. This is caused by the fact that GP in general prefer
the reconstructed functions to be smooth if there is no sufficient
evidence for rapid variations. The strength of this preference depends
on the choice of covariance function: the wider the peak of the covariance
function, the stronger the preference for a smooth reconstruction. 

We recommend the Mat\'ern($9/2$) covariance function for the
reconstruction of dark energy with current supernovae data as it provides reliable results for the
smoothest reconstructions that are consistent with the data. However,
keep in mind that it does not pick up rapid variations in $w$ if there
is insufficient evidence for these variations. In such a case, the
errors will be underestimated, and for a more conservative estimate one should use a smaller $\nu$. 

In general these results are indicative of a general trend for the reconstruction of $w$, but for substantially increased numbers of SNIa, very different errors or distributions, and different considerations in the families of EOS one might be interested in, this analysis should be extended. Once the numbers of SNIa becomes sufficiently large, more complicated covariance functions with different length scale parameters could also be considered for complicated EOS.  

We introduced a way to quantify deviations of a model from the GP
reconstructions. The coverage obtained for the actual data set needs
to be compared to the distribution of coverages of many mock data
sets. The input model for the mock data sets needs to be identical to
the model one wishes to test. The number of mock data sets with a
larger coverage than the one for the observed data determines the CL
at which the model can be rejected.

\acknowledgments

MS was supported by the South Africa Square Kilometre Array Project
and the South African National Research Foundation (NRF). CC was
supported by the NRF.

\bibliography{refs}

\end{document}